# The Impact of Oil and Gold Prices' Shock on Tehran Stock Exchange: A Copula Approach

**Amir T. Payandeh Najafabadi** [*]
Mathematical Sciences Department
Shahid Beheshti University
amirtpayandeh@sbu.ac.ir

**Marjan Qazvini**
E.C.O College of Insurance
Allameh Tabatabai University

**Reza Ofoghi**
E.C.O College of Insurance
Allameh Tabatabai University

## Abstract

There are several researches that deal with the behavior of SEs and their relationships with different economical factors. These range from papers dealing with this subject through econometrical procedures to statistical methods known as copula. This article considers the impact of oil and gold price on Tehran Stock Exchange market (TSE). Oil and gold are two factors that are essential for the economy of Iran and their price are determined in the global market. The model used in this study is ARIMA-Copula. We used data from January 1998 to January 2011 as training data to find the appropriate model. The cross validation of model is measured by data from January 2011 to June 2011. We conclude that: (**i**) there is no significant direct relationship between gold price and the TSE index, but the TSE is indirectly influenced by gold price through other factors such as oil; and (**ii**) the TSE is not independent of the volatility in oil price and Clayton copula can describe such dependence structure between TSE and the oil price. Based on the property of Clayton copula, which has lower tail dependency, as the oil price drops, stock index falls. This means that decrease in oil price has an adverse effect on Iranian economy.

## 1. Introduction

The Tehran Stock Exchange market (TSE) is the main capital market in Iran. In 2004, the capitalization of the TSE was 37.5 billion US$, which represents about 24 percent of GDP (the International Monetary Fund, say IMF, 2004). In 2008, the TSE displays a boom and became the most demanded investment market in Iran, particularly when the real estate started to decline. This fact along with privatization of public companies and high inflation in Iranian economy resulted in a dramatic growing trend in the TSE. At the same time, the price of two commodities that are vital for the economy of Iran kept rising in global market.

Oil plays a vital role in Iranian economy. Iranian annual budget relies heavily on oil export revenue. According to IMF (2010), in 2008/09, 74% of Iran's export was oil. This accounts for about 24% of GDP. Also, 65% of fiscal revenues, which is 16% of GDP was from oil. Therefore, one may expected increasing oil price may consequence booming in TSE.

On the other hand, Gold is a safe haven investment for all economies, particularly those that struggle with high inflation. A safe haven has relatively stable value (Bashiri, 2011). While national currency depreciates, the value of gold will not fall. This reason and the fact that housing return is greater than the rate of inflation (Masron and Fereidouni, 2010) makes Iranian people believe that *gold* and *real estate* are two important and profitable sources of investment given high inflation and currency depreciation in Iran- inflation was about 25% in 2008-(IMF, 2011). Thus, one can consider them as *competitors* for the TSE; since a change in gold price plays a significant role in people's encouragement or discouragement for investing in Stock Exchange market (SE).

The impacts of oil price on the SE (and via versa) have been studied by many authors, mostly on developed countries. Furstenberg and Bang (1989) observed that news about increasing oil price, negatively, impact on Japan and Germany SEs, but it positively impact on the UK and the US SEs. Park and Ratti (2007) verified same results for the long-run and the reverse for the short-run of 1 month in the US SE. El-Sharif et al. (2005), using a multi-factor regression model, observed that the stocks of energy sectors are affected, positively, by changes in oil price and negatively by the US$ exchange rate. Nandha and Hammoudeh (2007) studied the link between the performance of Asian-Pacific countries' SEs and global factors changes, such as oil price and foreign exchange rates. Using data collected over May 4, 1994 to June 30, 2004 for 15 regional countries, they found that: (i) Philippines and South Korea are sensitive to changes in oil price in the short-run whenever oil price is denominated in local currency instead of US$; (ii) Indonesia SE is responsive just to fall in oil price whenever it is expressed in local currency; and (iii) India, Indonesia, Malaysia, Singapore, Taiwan, and New Zealand show a significant relationship between the domestic stock index returns and changes in the exchange rates. Kilian and Park (2009) using a vector autoregressive model found that about 1% of variation in the US real stock returns could be explained by oil price shock in the short-run, whereas this effect rose to 22% in the long-run. Lin et al. (2009) found a positive oil 's influence on supply and demand shock of China and Hong Kong SEs, but no significant effect on Taiwan SE. Using a two factor regression model, Mohanty et al. (2010) found that, during December 1998 to March 2010, no evidence to support any relationship between equity values of oil and gas firms (in Hungary, Poland, Romania, Slovenia, and Czech Republic) and oil price. A negative relationship between oil price changes and stock returns in financial sector of European market, in short-term, has been observed by Arouri (2011). Using switching transition error correction model, Jawadi et al. (2010) found that there is a negative relationship between the oil price and French, the USA, and Mexican SEs. Constantinos et al. (2010) employed a vector autoregressive model along with a Granger-causality analysis and showed that there is a significant positive association between Greek SE and oil price.

A small number of studies have explored the effect of changes in gold price. For example, Zhang and Wei (2010) utilized the co-integration theory along with error correction model. They gathered daily data from January 4, 2000 to March 31, 2008 for oil Brent spot price (from the US Energy Information Agency (EIA) in US dollars), gold global prices (based on the London PM fix in US dollars) and examined the relationship between the crude oil and gold price. They observed that: (i) There is a high positive correlation between the crude oil price and the gold price, but the resulting effect of change in oil price lasts longer than that of gold price; and (ii) the change in crude oil price, causes the change in gold price.

There are some researches which investigate the responsiveness of the Middle East SEs to changes in oil price. Most of these studies analyze this effect on the GCC[1] SEs. In Oman, Qatar and the UAE the SEs positively react to oil price shocks. This relationship is nonlinear for Bahrain, Kuwait and Saudi Arabia (Basher and Sadorsky, 2006). Among the GCC countries, changes in the Saudi Arabia SE, which reflects changes in the Saudi Arabia economy leads to changes in OPEC oil price (Arouri and Rault, 2010). The same result has been found by Hammoudeh and Aleisa (2004). They concluded that there was a strong relation between Saudi Arabia SE and the NYMEX oil market, and political and economical situation in Saudi Arabia could reflect on oil future prices and predict it. In addition, shocks originating in the Saudi Arabia equity market affect the global oil market (Malik and Hammoudeh, 2007). Hammoudeh and Choi (2006) examined the co-movement among five GCC SEs and their links to three global factors- the WIT oil spot prices, the US 3-month Treasury bills rate and the SandP Index. Using weekly data collected from February 15, 1994 to December 28, 2004, they found that the US T-bill had a short term impact on some of the GCC markets. However, the oil price and the SandP500 index had no effect on GCC market in the short-run. A positive oil shock benefits most of GCC markets. Arouri et al. (2011), state that the rise in oil price volatility which is caused by shocks, and policy changes that affect oil supply and demand, would directly increase the volatility of the GCC markets.

It seems that Iran as a Middle East country, OPEC member, and one of the important oil producers in the world is short of such studies. There are a few studies that look at the effect of different macroeconomic variables on Iranian economy and the TSE. Farzanegan and Markwardt (2009) used a vector autoregressive model and data collected in post Iran-Iraq war period to analyze the relationship between oil price shocks and the Iranian economy. They found that: (i) The positive oil price shocks increase the real effective exchange rate; (ii) Both positive and negative demand shocks lead to higher inflation in Iranian economy; and (iii) There is a positive relationship between positive oil supply shock and industrial production growth and negative shocks reduce this growth. Abbasian et al.(2008) used a vector error correction model and quarterly data collected in period 1998 to 2005 to examine the effects of macroeconomic variables on the TSE. They found that: (i) Exchange rate and trade balance have a positive effect on the TSE in the long-run; and (ii) Liquidity, inflation and interest rate have a negative effect on the TSE.  Foster and Kharazi (2008) applied several statistical models, using weekly data collected from September 29, 1997 to November 18, 2002, to study the correlation between weekly returns of Tehran stock index and oil price. They found that oil price did not have significant impact on the TSE. Similar findings have been achieved by Samadi et al. (2007), but they concluded that gold price had an impact on the TSE. Zare and Rezaei (2006) used a vector error correction model to study how the TSE is influenced by foreign exchange, gold coin and housing markets. They found that the price of gold coin and real estate have positive effects on the TSE, while exchange rate has negative effect. Contrary to their findings, Bashiri (2011) found that there is no relationship between price of gold and stock index. She looked at the association between gold price and stock price index both in Iran over the period 1995 to 2010 and in Armenia over 2005 to 2010. She concluded that in contrast to developed countries that gold is an alternative to stock, there is relationship between price of gold and stock index neither for Iran nor for Armenia. Mashayekh  et. al. (2011), investigated the impact of inflation, interest rate and gold price on the TSE. They found a positive relation between inflation and bond yield on TSE. But the association between interest rate and TSE was negative. They concluded that there is no significant relationship between gold price and stock return. Saeidi and Amiri (2009) used a regression method and quarterly data from 2002 to 2008 to study the effect of macro variables such as: consumer price index (CPI), exchange rate, and oil price. They found that the CPI and exchange rate did not have any impact on the stock index, but there was a negative relation between oil price and stock index.

Keshavarz and Manavi (2009) employed a vector autoregressive model along with Granger causality test to investigate the impact of oil price volatility on the TSE and exchange rate using daily data collected from 27 March 1999 to 17 October 2006. The results showed that: (i) Increase in oil price affects the TSE from 1999 to 2000, but it does not have any effect on exchange rate from 2000 to 2001; (ii) During period 2001 to 2003 the TSE has an impact on the exchange rate and oil's impact on the TSE is insignificant; (iii) The oil price affects TSE and consequently the exchange rate within 2003-2004; (iv) In 2004-2006 there is no causal relationship, but it seems that exchange rate is more influenced by oil price and the TSE than other markets; and (v) In general, the TSE is more responsive to oil price increase. Impact of macroeconomic variables and competitor markets on the Iranian stock price index has been studied by Eslamlouian and Zare (2007). They employed an autoregressive distributed lag (ARDL) model with quarterly data collected from period 1993 to 2003 and introduced an appropriate Capital Asset Pricing Model (CAPM) for Iranian economy. They used Pesaran's approach to evaluate the ARDL model and concluded that the ratio of domestic to foreign price levels, housing, gold coin and oil price had positive impacts on the stock price, while exchange rate and money supply had negative impact, see: Pesaran and Shin, 1999 for more details on Pesaran's method. The effect of industrial outputs was insignificant.

Intuitively, there is a correlation among economical factors such as equity returns, oil price, exchange rate, etc. One way to show such correlation is the copula approach as an alternative to multivariate time series models. Jondeau and Rockinger (2006) utilized a time-varying parameter copula to study relationship between four major SEs pair wise: SandP500, FTSM, DAX, and CAC. Traditionally, such investigation has been conducted through either a time-varying correlation or a Markov-Switching model. They found that a time-varying parameter copula is a suitable model for the European markets. Roch and Alegra (2006) applied copula method to study the equity returns in Spanish SE. Ning (2010) employed the copula approach to study the link between equities and foreign exchange rates for G5 countries before and after Euro. Wang et al. (2011) showed that relationship between Chinese and the US market was well described by a Clayton copula, while a Gaussian copula explained the association between Chinese market and other markets such as Europe, Japan, and Pacific countries. Reboredo (2011) used a time-varying copula and a time-invariant copula with weekly collected data from 3 January 1997 to 4 June 2010, to study the association between different oil markets: WTI, Dubai, Maya, and Brent.

As mentioned above, there are different factors that are expected to impact the TSE such as interest rate, foreign exchange rate, gold, and oil price. In Iran, the interest rate is strictly controlled by the government and not the market (Mashayekh et. al., 2011). Statistics provided by the Central Bank of the Islamic Republic of Iran can endorse this fact. From 2001 to 2008, the short-term investment deposit rate was 7%. The change in US$ also is not pronounced and for many years remain relatively constant. However, gold and oil price are the only factors that are not determined domestically i.e. they are priced on the global market.

The purpose of this study is to investigate how the TSE is responsive to change in oil and gold price through applying the copula approach. This article is structured as follows: Section 2 discusses ARIMA and copula, which are two models used in this study. Sections 3 and 4 apply ARIMA and copula models to data collected from January 1998 to January 2011. In Section 5 the cross validation of the model is examined by data from January 2011 to June 2011 and the TSE will be predicted for the remaining six months. Finally, Section 6 concludes based upon the well fitted model.

## 2. Methodology

Now, we collect some useful elements which play central role in the next sections.

## ARIMA

The ARIMA processes are, in theory, the most popular time series model which can be stationarized by transformations such as differencing and logarithmic. In fact, the easiest way to think of ARIMA models is as fine-tuned versions of random-walk and random-trend models.

The model $\phi(L)\Delta_t^{dy} = \theta(L)a_t$ in which $\phi(L)\Delta^d = \varphi(L) = 1 -$

$\varphi_1 L - \cdots - \varphi_{p+d} L^{p+d}$, and $\theta(L) = 1 + \theta_1 L + \cdots + \theta_q L^q$ is called an autoregressive-integrated-moving average process of order $(p, d, q)$ and denoted as ARIMA (p, d, q), see: Harvey (1993, Section 3) for more details. An ARIMA model can be specified with the Box-Jenkins approach. If the series is not stationary and has trend, logarithmic transformation and differencing can make it stationary and remove the trend. The sample autocorrelation function (ACF) and sample partial autocorrelation function (PACF) are employed to determine the q and p, respectively. To examine the adequacy of the model, residuals must be analyzed. They should not form any pattern, and the p-value for Box-Pierce (Ljung-Box) statistic should accept the null hypothesis, which states that the residuals are independent and normally distributed.

**Copula**
There are different ways to show the relationship between random variables. One way is to use correlation coefficients. However, they can just describe the relationship between two random variables. Also, using these coefficients we cannot understand the nature of the relationship between random variables since they do not give any information about the variation of variables across their distributions. Papachristou (2004), with an interesting example shows that Pearson correlation coefficient alone cannot give useful information about a set of data, therefore he along with other researchers suggested to use copula to define the dependency among random variables. He noted that it is possible to have two distributions with the same correlation, but different dependence structure. In this context, correlation coefficient is similar to the mean that gives only limited information about the distribution function.

Copula is a multivariate dependence function; as its name implies it links the marginal distributions of random variables to their joint distribution function. Each copula can be joined with different marginal distribution functions to form a joint multivariate distribution function. There is no such flexibility for other multivariate distributions, say, multivariate normal distributions. With multivariate normal distributions we are bound to have normal distributions for all margins. Another important property of copula is that it is *invariant* under strictly increasing transformation. Copula has route in different sciences, such as actuarial sciences: in credibility, and modeling losses with expenses associated with them (Frees and Valdez, 1998; Frees and Wang, 2005); finance: in risk management, asset allocation and derivative pricing (Cherubini et al., 2004); hydrology (Genest and Favre, 2007); biological studies: in epidemiology (Clayton, 1978). The cornerstone of copula is based on a theorem proposed by Sklar in 1959, see Cherubini et al. (2004) for more detail.

According to Sklar's theorem if $X = (X_1, \cdots, X_d)$ is a continuous random vector with continuous distribution functions $F_1(\cdot), \cdots, F_d(\cdot)$ and joint cumulative distribution function $F_X(\cdots)$ then there exists a unique copula $C_\theta$ which is defined on $[0,1]^d$ with uniform margins such that for all $x \in \Re^d: F_X(t_1, \cdots, t_d) = C_\theta(F_1(t_1), \cdots, F_d(t_d))$, where $\theta$ stands for parameter of copula, for more information on copula see: Nelsen (2006), Denuit et al.(2005), among so many others.

Elliptical and Archimedean copulas are two important classes of copulas. Gaussian and t-copulas are two types of elliptical copulas, which are symmetric and extremely used in finance; however t-copula is normally a better choice, because apart from being symmetric, it is possible to observe more dispersion in its tails. Frank, Clayton and Gumbel are three important Archimedean copulas. Frank is a copula that shows neither upper nor lower tail dependence. Gumbel only exhibits upper tail dependence and Clayton lower tail dependence. Tail dependence measures the dependence structure of extreme values. Kendall's tau and Spearman's rho, which are rank-based measures, also show dependency among variables. Table 1 demonstrates the relation between Kendall's tau, $\tau$, and parameters of some copulas.

**Table 1: Two dimensional Copulas, the relations between parameters and Kendall's tau, $\tau$.**

| Copula | Functional form | Parameter |
|--------|-----------------|-----------|
| Gaussian | $C_R^{Ga}(u_1, u_2) = \Phi\left(\Phi^{-1}(u_1), \Phi^{-1}(u_2)\right)$ | $R = sin(\pi\tau/2)$ |
| T ($df=v$) | $C_R^t(u_1, u_2) = t_{v,R}\left(t_v^{-1}(u_1), t_v^{-1}(u_2)\right)$ | $R = sin(\pi\tau/2)$ |
| Gumbel | $C_\theta^{Gumbel}(u_1, u_2) = exp\left(-\left[(-ln\, u_1)^\theta + (-ln\, u_2)^\theta\right]^{\frac{1}{\theta}}\right)$ | $\theta = 1/(1-\tau)$ |
| Clayton | $C_\theta^{Clayton}(u_1, u_2) = \left(u_1^{-\theta} + u_2^{-\theta} - 1\right)^{-/\theta}$ | $\theta = 2/(\tau(1-\tau))$ |

To select the right copula among different classes, one has to apply goodness-of-fit tests. The null hypothesis is $H_0: C \in C_0 = \{C_\theta: \theta \in O\}$, where $O$ is an open subset of $\mathfrak{R}^d$ for integer $d \geq 1$. In a review study by Genest et al.(2009) some so-called ``blanket tests'' (such as Kolmogorov-Smirnov, Cramér-von-Mises, Anderson-Darling, Rosenblatt transform, etc.), which are free of any subjective decisions regarding the form of, say, kernel estimation, bandwidth, etc. are provided. The most powerful test which is also used in copula R Package (R Development Core Team, 2011 and Kojadinovic and Yan, 2010) is the Cramér-von-Mises:

$$S_n = \int_{[0,1]^d} C_n(u)^2 \, dC_n(u) = \sum_{i=1}^n \{ C_n(\hat{U}_i) - C_{\theta_n}(\hat{U}_i) \}^2, \qquad (1)$$

where $C_n(u) = \sum_{i=1}^n 1\{\hat{U}_i \leq u\}$, $u \in [0,1]^d$. The Cramér-von-Mises statistic is an empirical distribution function, namely, EDF type statistic, which is based on discrepancy between the hypothesized distribution function $C_\theta$ and the empirical distribution function $C_n$. The smallest $S_n$ will lead to selection of the best copula. Moreover, the basis of the Kolmogorov-Smirnov test is a measure of discrepancy between the hypothesized distribution function $F_0$ and the EDF $\widehat{F_n}$ see: Lehmann and Romano (2005, Section 14) for more details.

**ARIMA-Copula**
Interaction among several time-dependent random variables can be studied using either the *multivariate time series* or the *ARIMA-Copula* modeling. The multivariate time series approach needs a deep understanding of the problem in hand. Moreover, only a few statistical packages can handle such approach (Brockwell and Davis, 2009, Section 11). A time-dependent random variable can be decomposed into a time series model and residuals, where residuals are time-independent (Brockwell and Davis, 2009, Section 1). The ARIMA-Copula modeling is simply benefit from the above fact and considers the impact of several time-dependent random variables on each other, in two stages: (**i**) Find an appropriate ARIMA model for each time-dependent random variable; and (**ii**) Find an appropriate time-invariant Copula for residuals of all fitted ARIMA models. The validity of fitted Copula can be evaluated since it is a time-invariant Copula. Now, one can study interaction among time-dependent random variables through such fitted Copula model, see Melo Mendes and Ai'ube (2011), Grégoire et al. (2008), among others, for other applications of the ARIMA-Copula model.

**Cross validation**

After building the model we need to examine its validation. Cross validation is a method in which the data are split into two sets: (i) Training set and; (ii) Testing set. The former is used to establish the model and the latter to verify its validation (Hastie et al., 2009, Section 7). In Section 5 this method will be applied to the model and MSE will be used as a criterion to show the performance of the model.

### 3. Do oil and gold prices impact on the TSE?

To answer the above question, the data for Tehran Stock Index, oil and gold price over the period January 1998 to January 2011 have been collected. The data are available at www.indexmundi.com and www.tse.ir. The data for crude oil Brent and gold price are denominated in US$. The index used for the TSE is actually the closing price. The monthly average of daily data is used to avoid any problem that may arise due to difference between Iranian and Christian holidays and also to avoid too much random fluctuation and white noise that may apparent in daily data. Monthly data are spread, therefore fluctuation in the time series will be smoother, (Mun, 2006). The data for TSE are converted from Persian to Western months. The log-transformed data are used in order to reduce the big variations of data. The time series plot of the log-transformed data is illustrated in Figure 1.

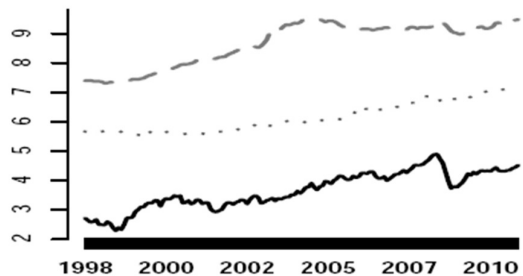

**Figure 1: This figure shows the co-movement of log-transformed price of oil (solid line), gold (dot line), and stock index (dash line)**
(Source of Data: http://www.indexmundi.com/and http://www.tse.ir/).

As one can observe from Figure 1, all three variables are increasing over the years. More volatility can be spotted for the log-price of oil. As one can see, during 1998 the oil price goes up and down until it falls to 9.8$ in Dec 1998 which is the lowest value for oil over the period under observation. It seems that the oil price as most other economical factors exhibits a cyclic pattern. By looking at stock trend, one can detect two peaks for the stock index: (i) In Dec 2004 the value was 13,687; (ii) It was $12,640$ in July 2008. The last one is concurrent with the maximum rise in oil price which happened in 2008. As of 2008 the behavior of stock index is more in agreement with the oil and gold price and accordingly to the global market. Late 2008, coincides with the outbreak of financial crisis and presidential election in the US which can well explain these price changes.

The objective of this research is to understand the structure of the impact of oil and gold price on the TSE and use the model to forecast the stock index. To achieve this purpose, first the marginal distributions and then the copula class must be specified.

**Marginal distributions**

As mentioned above, in the first step, one has to fit *either* marginal distribution *or* marginal time series model to each variable in hand.

The Box-Pierce (Ljung-Box) test given in Table 2 shows that the null hypothesis $H_0$: *data are time independent* must be rejected, at 5% level, in favor of $H_1$: *data are time dependent*. Therefore the data must be analyzed using a time series approach.

**Table 2: the Box-Pierce (Ljung-Box) test**

|  | p-value | x-squared |
|---|---|---|
| Log-stock | $2.2 \times 10^{-16}$ | 151.3536 |
| Log-oil | $2.2 \times 10^{-16}$ | 148.1719 |
| Log-gold | $2.2 \times 10^{-16}$ | 149.6905 |

Following the Box-Jenkins technique, the models found for log-stock, log-oil and log-gold are ARIMA(0,2,2), ARIMA$(1,1,0)(1,0,1)_{11}$ and ARIMA (0,2,1), respectively. The following table represents coefficients, SE, t-value and p-value for null hypothesis that the corresponding coefficient is zero.

**Table 3: The fitted ARIMA models.**

|       |     |    | Coefficient | SE     | t-value  | p-value |
|-------|-----|----|-------------|--------|----------|---------|
| Stock | MA  | 1  | 0.2565      | 0.0624 | 4.11     | 0.000   |
|       | MA  | 2  | 0.6380      | 0.0625 | 10.22    | 0.000   |
| Oil   | AR  | 1  | 0.2061      | 0.0800 | 2.58     | 0.011   |
|       | SAR | 11 | 0.8768      | 0.1260 | 6.96     | 0.000   |
|       | SMA | 11 | 0.7346      | 0.1738 | 4.23     | 0.000   |
| Gold  | MA  | 1  | 0.9999      | 0.0000 | 28210.60 | 0.000   |

The results from Table 3 show that coefficients of all ARIMA models are significant, at significant level $\alpha = 0.05$. To verify the appropriateness of the fitted ARIMA model, one has to study residuals of such models.

**Table 4: The Shapiro-Wilk normality test.**

| Residuals of | W statistic | p-value |
|--------------|-------------|---------|
| Oil          | 0.9802      | 0.02489 |
| Gold         | 0.9773      | 0.01203 |
| Stock        | 0.9834      | 0.06141 |

The Shapiro-Wilk normality test, given in Table 4, suggested that the residuals are normal at significant level $\alpha = 0.01$. Time independency of residuals has been verified by Figure 2 and Table 5.

**Table 5: The Box-Pierce test results for residuals of fitted models**

| Time series model | p-value |
|-------------------|---------|
| Stock             | 0.6062  |
| Oil               | 0.8478  |
| Gold              | 0.6233  |

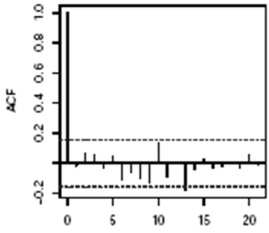

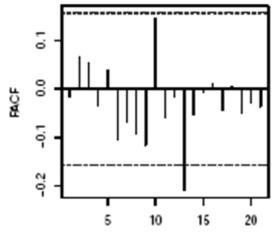

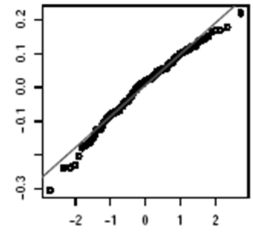

(a)

(b)

(c)

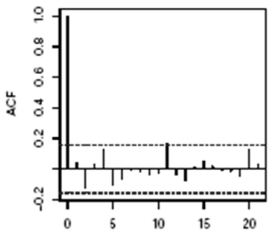

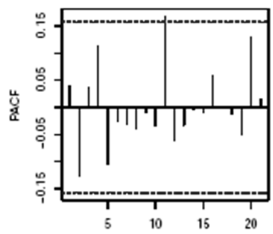

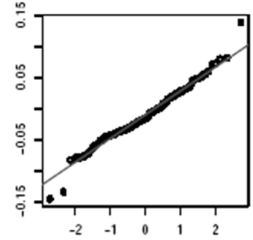

(d)

(e)

(f)

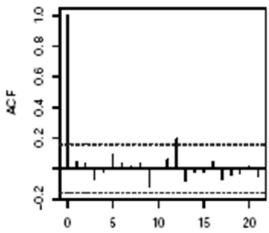

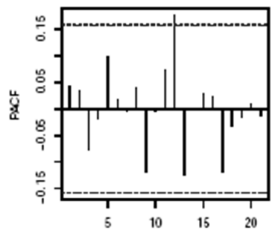

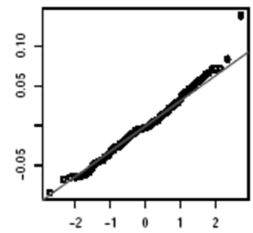

(g)

(h)

(i)

**Figure 2: Figures (a) to (i) represent the ACF, the PACF, and the quantiles plot of residuals of three time series models: Oil, Gold, and Stock, respectively.**

The results of Tables 3, 4, and 5 along with Figure 2 verified the appropriateness of the fitted ARIMA models. The fitted ARIMA models for logarithm of oil, stock and gold are, respectively, given bellow.

$$Y_t^{Stock} = 2\,Y_{t-1} - Y_{t-2} + a_t - 0.0624\,a_{t-1} - 0.0625a_{t-2} + error \tag{2}$$

$$Y_t^{Oil} = 1.206\,Y_{t-1} - 0.2061\,Y_{t-2} + a_t + error \tag{3}$$

$$Y_t^{Gold} = 2\,Y_{t-1} + Y_{t-2} + a_t - 0.999a_{t-1} + error \tag{4}$$

The seasonal pattern (SARIMA), lag 11, for oil price may be justified by the facts that: (i) The equity market in Saudi Arabia affects the global oil price (Malik and Hammoudeh, 2007, Arouri and Rault, 2010, and Hammoudeh and Aleisa, 2004); and (ii) Saudi Arabia is an Islamic country, where ``Ramadan'' is a sacred month in which Muslims spend their times for religious affairs and pray rather than business and economic activities. Therefore, the lag of 11 can be interpreted in a way that there are 11 months in a year that economy is fully active, rather than 12 months.

**Copula**

This study deals with time-invariant copulas. Therefore, one has to work with the residuals which are time-independent. This section aims at finding a well-fitted copula for residuals of three above ARIMA models. To do so, one has to measure the correlation between the TSE, oil and gold prices using either the Kendall's tau or Spearman's rho. Pearson correlation coefficient can be used as well, though it is not a concordance measure as Kendall's tau and Spearman's rho. The two random variables are said to be concordant if large values of one tend to be associated with large values of the other and small values of one with small values of the other see: (Nelsen, Section 2006).

**Table 6: The Kendall's tau and Spearman's rho[*] measures for residuals with their (p-values).**

|        | Stock           | Oil                              | Gold                             |
|--------|-----------------|----------------------------------|----------------------------------|
| Stock  | 1               | $0.138^*$ (0.0438)               | 0.039 (0.2347)                   |
| Oil    | 0.094 (0.04215) | 1                                | 0.688 ($2.2 \times 10^{-16}$)    |
| Gold   | $0.068^*$ (0.2018) | $0.877^*$ ($2.2 \times 10^{-16}$) | 1                                |

Such nonparametric correlations, represented in Table 6, show that there is significant correlation between the TSE and oil price at 5% significant level; while, correlation between the TSE and gold price is insignificant at 5% significant level. This means that the product copula i.e. $C(u, v) = uv$, is an appropriate dependence function to represent dependence structure between gold price and the TSE. However, the association measures between oil and gold price are significant which means that gold price can impact, indirectly, on the TSE through its impact on oil price. This makes sense, when we apply the multivariate independence test (Kojadinovic and Yan, 2010) to our three variables and the null hypothesis of independence is rejected with p-value= 0.00249. Therefore, although the TSE and gold price are pair wise independent, our three variables are dependent altogether.

To specify the copula that defines the nature of the relationship between oil price and the TSE, one has to apply the goodness-of-fit test for Clayton, Frank, Gumbel, Plackett, normal, and t copulas. The goodness-of-fit tests of such class of copulas are presented in Table 7.

**Table 7: The accepted copulas which represent relationship between stock index and oil price**

| Copula | Parameter | Crame r-von-Mises statistic | p-value |
|--------|-----------|-----------------------------|---------|
| t (df=25) | 0.152870 | 0.022595 | 0.3661339 |
| Normal | 0.154899 | 0.022562 | 0.3851149 |
| Frank | 0.872933 | 0.021595 | 0.3931069 |
| Clayton | 0.303698 | 0.018178 | 0.6418581 |
| Plackett | 1.578407 | 0.020530 | 0.514985 |

According to p-values, which are greater than 5% all copulas but Gumbel pass the goodness-of-fit test. However, one may select the well-fitted copula based on the higher p-value and lower Crame r-von-Mises statistic. It seems that Clayton copula with p-value = 64% can perform better than other copulas. The Clayton copula as mentioned above demonstrates lower tail dependence. Based upon the well fitted ARIMA model for log-stock and a Clayton copula which represents dependence structure between residuals of log-stock and log-oil ARIMA models, one can conclude the following well fitted ARIMA model along with a Clayton copula with parameter $\theta = 0.303698$, say, ARIMA-Clayton model.

$$Y_t^{Stock} = 2\,Y_{t-1} - Y_{t-2} + a_t - 0.0624\,a_{t-1} - 0.0625a_{t-2} + \psi^{-1}(r), \qquad (5)$$

where $\psi(x) = \int_{-\infty}^{x}\int_{-\infty}^{\infty}\frac{\partial^2}{\partial t_1 \partial t_2}\left[\Phi\left(\frac{t_1}{0.035}\right)^{-\theta} + \Phi\left(\frac{t_2}{0.094}\right)^{-\theta} - 1\right]^{-\frac{1}{\theta}} dt_1 dt_2$, stands for residuals for ARIMA models (Equations 2 to 4) and $\Phi(\cdot)$ stands for cumulative distribution function of standard Normal distribution.

Using the ARIMA-Clayton model (Equation 5), and based on the property of Clayton copula one may deduce that decrease in price of oil is accompanied by decrease in stock index. In other words, stock index and oil price fall together when economy is in downturn. However, it is not possible to come up with the same conclusion just by looking at Figure 1.

**Cross validation and forecasting**

There are not many researches on forecasting with copula. Grégoire et al. (2008) is one of the article which deals with forecasting with Copula . In that article, they looked at the relationship between oil and gas price and predicted the price of gas via copula. This study, in the same fashion with Grégoire et al.(2008), predicts the log-stock index for 12 months ahead, i.e., from January 2011 to December 2011. For this purpose one has to employ the ARIMA-Clayton model (Equation 5).

The cross validation of the model is measured by the first 6 months. Table 8 employed the fitted ARIMA models (Equation 2 to 4) and ARIMA-Clayton model (Equation 5) to predict stock price for 12 months of 2011.

**Table 8: Stock predicted values for 12 months of 2011 using Equations 2 to 5**

| Month | Stock Prediction using | | Log-Actual Stock price |
|---|---|---|---|
| | The fitted ARIMA model (Equation 1) | ARIMA-Clayton model (Equation 2) | |
| Jan. | 9.506 | 9.513 | 9.538 |
| Feb. | 9.519 | 9.522 | 9.395 |
| March | 9.533 | 9.526 | 9.699 |
| April | 9.546 | 9.549 | 9.825 |
| May | 9.560 | 9.561 | 9.763 |
| June | 9.574 | 9.578 | 9.706 |
| July | 9.588 | 9.592 | -- |

| Month | Stock Prediction using | | Log-Actual Stock price |
| | The fitted ARIMA model (Equation 1) | ARIMA-Clayton model (Equation 2) | |
| --- | --- | --- | --- |
| August | 9.601 | 9.613 | -- |
| September | 9.615 | 9.621 | -- |
| October | 9.628 | 9.630 | -- |
| November | 9.642 | 9.649 | -- |
| December | 9.656 | 9.650 | -- |

Comparing the first six month actual and predicted values, given in Table 8, one may conclude that the ARIMA-Clayton model (Equation 5) provides more accurate prediction. Figure 3 also illustrates such observation.

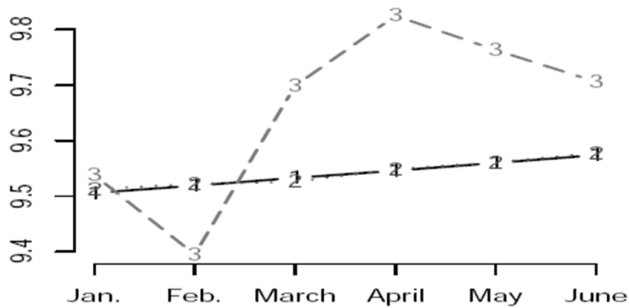

**Figure 3: dash line numbered 1 to 3 represent, respectively, actual log-stock index; predicted log-stock index using the ARIMA models (Equation 2 to 5); and predicted log-stock index using the fitted the ARIMA-Clayton model (Equation 5).**

To verify the above observation more precisely, one may define the following mean-squared error (MSE).

$$MSE = \frac{1}{n}\sum_{t=1}^{n}\left(Y_t^{Stock} - \hat{Y}_t^{Stock}\right)^2. \qquad (6)$$

As explained in Methodology, the MSE can be viewed as a criterion which measures accuracy of the ARIMA and ARIMA-Clayton model (Equation 5) in prediction of stock prices.

The MSE for the ARIMA model is 0.03007 while the MSE for the ARIMA-Clayton model (Equation 5) is 0.030007. This observation confirms the above claim that ``The ARIMA-Clayton models (Equation 2 to 4) provides more acceptable and accurate prediction for stock index compared to the ARIMA model". Using the ARIMA-Clayton model (Equation 5), the following figure illustrates prediction density functions for the next 6 months of 2011, i.e., form July 2011 to December 2011.

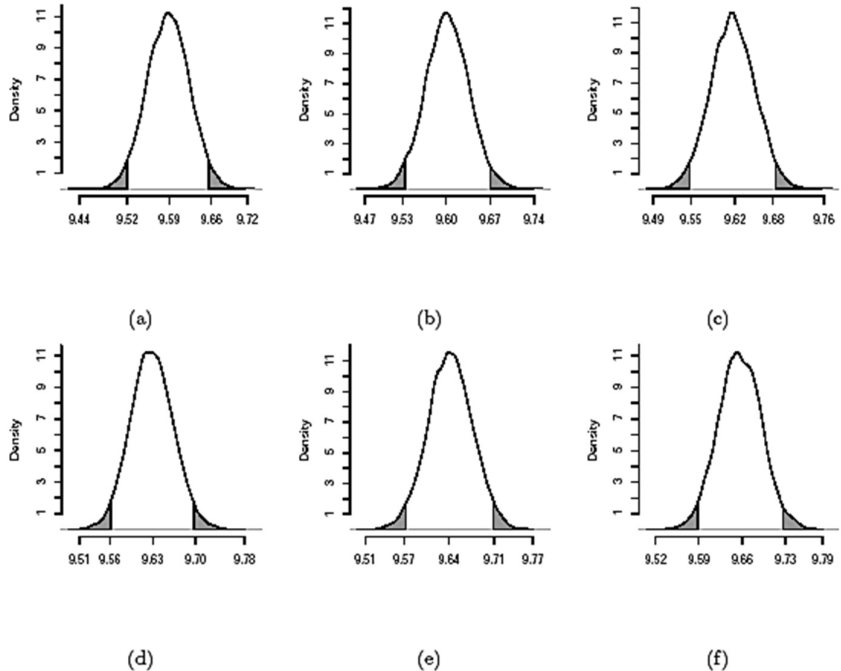

**Figure 4: Figures (a) to (f), respectively, represent prediction density function with 0.95% convinced interval for log-stock index from July 2011 to December 2011.**

From Figure 4, one may observe that with 95% confidence the log-stock index falls within [9.52, 9.66], in July and it slowly changes to [9.59, 9.73], in December. Also as the density for stock index is fairly normal, the median can represent the predicted value for stock index in forthcoming months. This prediction can be viewed as being conditioned on any values of oil and gold price i.e. each variable is predicted independent of others.

## 4. Conclusions

This article investigates the impact of oil and gold prices on the TSE.

It uses the log-transformed data and fits an ARIMA process to each series. The ARIMA for oil was seasonal (SARIMA) with lag 11. Referring to the literature, we ascribed the result to the economic activities in Saudi Arabia and the fact that the economy is not moving in "Ramadan", therefore it is possible to see a year with 11 active months. The residuals for each series were time independent and suitable for invariant-parameter copulas. However, the test for independency showed that there was no significant direct relationship between stock index and the gold price, which satisfies the result found by Bashiri (2011), among others. But test for multivariate independency indicated that it could be affected by gold price, indirectly. The appropriate copula for stock index and oil price turned out to be Clayton copula, which has lower tail dependency; therefore we can conclude that a fall in stock index comes with a reduction in oil price. This implies that as Iran relies on the revenue from oil export, decrease in oil price has an adverse effect on its main capital market, and consequently the economy. This observation also verified by several authors such as Farzanegan and Markwardt (2009), among others. We used the Clayton copula to forecast the stock index for 12 months of 2011. The cross validation of the model was verified by data from the first 6 months of 2011. Finally using the ARIMA-Clayton model (Equation 2) density functions of the next 6 months of 2011 have been predicted.

At the time of writing this article the government's monetary policy has changed and interest rate were reduced considerably. This fact along with some reports about the change of currency made stock investment an appealing option and economists warned about the bubble that started to form on the TSE. This study did not allow for such economical changes. In Iran interest and foreign exchange rate are controlled by the government and are not influenced by market. That is why they are excluded from our model. Recent change in monetary policy can attest such claim. Some researchers raised the issue of the role of Saudi Arabia in global oil price and our finding satisfies their results, it seems that there is a relationship between Saudi and Iranian economy at least through oil. Therefore, this issue can be the subject of future researches.

## Endnote

1. The Gulf Cooperation Council (GCC) was established in 1981.It includes six countries: Bahrain, Oman, Kuwait, Qatar, Saudi Arabia and the United Arab Emirates (UAE).